\documentclass[11pt,times]{article}

\usepackage{amssymb}
\usepackage{graphics,color}      % usual driver
\usepackage{verbatim}
\usepackage{epsfig}
\usepackage{amsmath}    % need for subequations
\usepackage{amsfonts}
\usepackage{times}
\usepackage{amsthm}

\newenvironment{prevproof}[2]{\noindent {\em {Proof of {#1}~\ref{#2}:}}}{$\Box$\vskip \belowdisplayskip}

\newcommand{\R}{\mathbb{R}}

\newcommand{\N}{\mathbb{N}}

\newcommand{\rmax}{\text{rmax}}

\newtheorem*{theorem1}{Theorem}
\newtheorem*{prop1}{Proposition}
\newtheorem{theorem}                {Theorem}[section]

\newtheorem{fact}       [theorem]   {Fact}

\newtheorem{definition}     [theorem] {Definition}

\newtheorem{proposition}       [theorem]   {Proposition}
\newtheorem{remark}        [theorem] {Remark}

\def\ud{{\mathcal{U}_D}}

\let\oldbibliography\thebibliography
\renewcommand{\thebibliography}[1]{%
  \oldbibliography{#1}%
  \setlength{\itemsep}{0pt}%
  \setlength{\parskip}{2pt}%
}

\begin{document}

\title{Size Bounds for Conjunctive Queries with General Functional Dependencies}
\author{Gregory Valiant and Paul Valiant\\\\University of California, Berkeley}

\maketitle
\begin{abstract}
This paper extends the work of Gottlob, Lee, and Valiant (PODS 2009)~\cite{GLV}, and considers worst-case bounds for the size of the result $Q(D)$ of a conjunctive query $Q$ to a database $D$ given an arbitrary set of functional dependencies. The bounds in~\cite{GLV} are based on a ``coloring'' of the query variables.  In order to extend the previous bounds to the setting of arbitrary functional dependencies, we leverage tools from information theory to formalize the original intuition that each color used represents some possible entropy of that variable, and bound the maximum possible size increase via a linear program that seeks to maximize how much more entropy is in the result of the query than the input.  This new view allows us to precisely characterize the entropy structure of worst-case instances for conjunctive queries with simple functional dependencies (keys), providing new insights into the results of~\cite{GLV}.  We extend these results to the case of general functional dependencies, providing upper and lower bounds on the worst-case size increase.  We identify the fundamental connection between the gap in these bounds and a central open question in information theory.

Finally, we show that, while both the upper and lower bounds are given by exponentially large linear programs, one can distinguish in polynomial time whether the result of a query with an arbitrary set of functional dependencies can be any larger than the input database. \end{abstract}

\section{Introduction}
In this paper, we are concerned with deriving worst-case size bounds for the result of a conjunctive query in terms of the structural properties of the query, and those of the input relations.  This paper addresses the main open question left by Gottlob, Lee, and Valiant (PODS 2009)~\cite{GLV}, extending size bounds to the case where the query is applied to a database that has an arbitrary set of general functional dependencies (as opposed to just `simple' functional dependencies---those whose left-hand sides consist of a single variable---as was done in~\cite{GLV}).

Conjunctive queries are the most fundamental and most widely used
database queries, forming the core of relational algebra~\cite{chandraMerlin,LMS,abitebook}.
Conjunctive queries also correspond to nonrecursive datalog rules
of the form
\begin{equation*}
%\label{($*$)}
R_0(u_0) \leftarrow R_1(u_1) \wedge \ldots \wedge R_n(u_m),
\end{equation*}
 where
$R_i$ is a relation name of the underlying database $D$,
$R_0$ is the output relation, and where
each argument $u_i$ is a list of $|u_i|$ variables,  where $|u_i|$ is the arity of
the corresponding relation, and where the same variable can occur
multiple times in one or more argument lists.
We allow a single relation $R_i$ to
appear several times in the query, thus $m \ge n$. Throughout this paper we
adopt this datalog rule representation for conjunctive queries.

In general,
the result of a conjunctive query can be
exponentially large in the input size. Even in the case of bounded
arities, the result can be substantially larger than the input
relations. In the worst case, the output size is
$r^k$, where $r$ is the size of the largest input relation and $k$ is
the arity of the output relation. Queries with very large outputs are
sometimes unavoidable, but in most cases they are either ill-posed or
anyway undesirable, as they can be disruptive to a multi-user DBMS.
It is thus useful to recognize such queries, whenever possible.
Obtaining good worst-case bounds for conjunctive queries is, moreover,
relevant to view management~\cite{LMS} and  data integration~\cite{lenzerini,LMS}, as
well as to data exchange~\cite{fagin03data,kola05schema}, where data is transferred
from a source database to a target database according to schema mappings
that are specified via conjunctive queries. In this latter context,
good bounds on the result size of a conjunctive query may be used for
estimating the amount of data that needs to be materialized at the
target site.

In the area of query optimization, models for predicting the size of the
output of a conjunctive query  based on selectivity indices for
relational operators have been developed~\cite{swami94,jarke-koch,chaudhuri-qo}. The selectivity
indices are obtained via sampling techniques (see, e.g.~\cite{OlkenR90,Haas}) from
existing database instances. Worst case bounds may  be obtained by
setting each selectivity index to 1, thus assuming the maximum
selectivity for each operator. Unfortunately, the resulting bounds
are then often trivial (akin to the above $r^k$ bound).

A new and very interesting
characterization of  the worst-case output size of {\em join queries}
was very recently developed by Atserias, Grohe, and Marx~\cite{AGM}. Their result is
based on the notion of {\em fractional edge cover}~\cite{MM}, and the
associated concept of  {\em fractional edge-cover number} $\rho^*(Q)$
of a join query $Q$. In particular, in~\cite{MM} it was shown
that \begin{equation}
~\label{hello}
|Q(D)|\leq \rmax(Q,D)^{\rho^*(Q)},
\end{equation}
where $\rmax(Q,D)$ represents the size of the largest relation among $R_1,\ldots,R_n$ in $D$.  In~\cite{AGM} it was shown that this bound is essentially tight.

In~\cite{GLV}, these results were extended beyond join-queries, to general conjunctive queries (containing projections) and also to the setting in which the input relations satisfy simple functional dependencies. This work introduced a new coloring scheme for query variables, and, accordingly,
the association of  a {\em color number} $C(Q)$ with each query $Q$.
Roughly, a valid coloring assigns a set $\mathcal{L}(X)$ of colors to
each query variable $X$ and requires that for each functional
dependency $X Y \rightarrow Z$, the colors of $Z$ are contained in the union of the colors of $X$ and $Y$. The {\em color number} $C(Q)$ of $Q$ is the maximum over all
valid colorings of $Q$ of the quotient of the number of colors appearing in the
output (i.e., head) variables of $Q$
by the maximum number of colors appearing in
the variables of any input (i.e., body) atom of $Q$.  It was shown that for a query $Q$ and database $D$ with a set of simple functional dependencies, $$|Q(D)|\leq \rmax(Q,D)^{C(Q)}.$$

In this paper, we attempt to extend these results to the case where we have a general set of functional dependencies (including compound functional dependencies of the form $X,Y,Z \rightarrow W$.)
In this setting, while the lower bound given by the color number holds, we illustrate that the color number no longer provides an upper bound on the worst-case size increase.  In fact, we provide a family of instances demonstrating that there is a super-constant gap between the true size increase and the bound given by the color number.

In order to provide size bounds in this general setting we require machinery beyond the color number.  We use tools from information theory developed to analyze the precise interactions of multivariate distributions.  In some sense, this approach formalizes the original intuition of the coloring scheme---that each color used represents some possible entropy of that variable.  We construct a linear program with entropies as the variables and the exponent of the worst-case size increase as the solution.  Functional dependencies can be encoded as constraints in the linear programs.  The difficulty is determining which additional constraints must be added to the linear program to ensure that the solution is realizable as a database instance.

This question, as it turns out, is crucially related to an old and ongoing investigation at the heart of information theory: ``which entropy structures can be instantiated in multivariate distributions?'' \cite{pip86,yeung97,yeung98,matus07,matus_inf,doug07}.  We cannot show that our upper bound is tight in this general setting, and believe that an explicit (even exponential-sized) characterization of the worst-case size increase is unlikely without significant advances in information theory.

Nevertheless, the formalism and tools from information theory shed significant light on the setting in which all functional dependencies are simple---the case considered in~\cite{GLV}.   We  revisit the color number, and the tight bounds on the size increase for queries with simple functional dependencies, providing an alternative formulation of the color number as the solution to a linear program whose variables are entropies.  This formulation allows us to show that the settings for which we have tight bounds on the size increase have worst-case instances with particularly simple entropy-structures; specifically, all associated mutual information measures are nonnegative.

Finally, while both our upper and lower bounds are given by linear programs that have exponentially many variables, we show that we can decide in polynomial time whether a query and set of functional dependencies is sparsity-preserving.  In particular, we can efficiently decide whether the result of a query can be any larger than the input database.

This paper is organized as follows. In Section~\ref{sec:prelim} we
state some useful definitions of database terms, define the coloring scheme and the color number of a query, and provide definitions of the basic information theory quantities and the Shannon information inequalities.  In  Section~\ref{sec:size} we identify the connection between entropy and worst-case instances, and prove our linear programming size bound.  In Section~\ref{sec:alt} we provide an alternative definition of the color number in terms of entropies, and identify the simple entropy structure of worst-case instances in the settings in which we have tight size bounds (the setting with simple functional dependencies).  We leverage this understanding of the entropy structure of these instances to construct a family of instances that demonstrate a super-constant gap between our upper and lower bounds.  Finally, in Section~\ref{sec:complexity}, we show that we can efficiently decide whether a query and set of functional dependencies can admit any size increase.

\section{Preliminaries}\label{sec:prelim}

We begin by giving basic definitions pertaining to database theory. We then define the color number, and state the size bounds of~\cite{GLV}. Finally, we define some information theoretic quantities, and define the Shannon information inequalities.

\subsection{Database Terminology}

As already stated in the Introduction,
a \emph{conjunctive query} has the form
$R(u_0) \leftarrow R_1(u_1) \wedge \ldots \wedge R_n(u_m),$ where
each $u_i$ is a list of (not necessarily distinct) variables of length $|u_i|=arity(R_i)$. Each variable occurring in the
query head $R_0(u_0)$ must also occur in the body of the query.
The set of all variables occurring in $Q$ is denoted by $var(Q)$.
It is important to recall that a single relation $R_i$
might appear several times in the query, and thus $m$ could be larger
than  $n$. A {\em finite structure} or {\em database}
$D=(\ud,R_1,\ldots,R_k)$ consists of a finite universe $\ud$ and
relations $R_1,\ldots,R_k$ over $\ud$. The answer $Q(D)$ of query $Q$
over database $D$ consists of the structure $(\ud,R_0)$ whose unique
relation $R_0$ contains precisely all tuples $\theta(u_0)$ such that
$\theta:var(Q)\rightarrow \ud$ is a substitution such that
for each atom $R_i(u_j)$ appearing in the query body,
$\theta(u_j)\in R_i$.  For ease of notation, we define $\rmax(Q,D)$ to be the number of tuples in the largest relation among $R_1,\ldots,R_n$ in $D$.

A {\em (simple) attribute} of a relation $R$ identifies a column of $R$.
An {\em attribute list}  consists of a list (without
repetition) of attributes of a relation $R$. A {\em compound attribute}
is an attribute list with at least two attributes.
A list consisting of a unique  attribute $A$ is identified
with $A$. The list of all attributes of $R$ is denoted by $attr(R)$.
If $V$ is a list of attributes of $R$ and $t\in R$ a tuple of $R$, then
the $V$-value of $t$, denoted by $t[V]$ consists of the tuple
obtained as the ordered list of all values in $V$-positions of $t$.

If $V$ and $W$ are (possibly compound) attributes of $R$, then
a {\em functional dependency (FD)} $V\rightarrow W$ on relation $R$ expresses that
for each $t,t'\in R$, $t[V]=t'[V]$ implies that $t[W]=t'[W]$.
Thus each functional dependency $V\rightarrow W$ is equivalent to
a set containing a FD $V\rightarrow A$ for each element $A$ of $W$.
If $A$ and $B$ are single attributes, then the FD $A\rightarrow B$ is
called a {\em simple FD}. A (possibly compound) attribute $K$ of $R$ is a
{\em key}
iff $K\rightarrow attr(R)$ holds.
Such a key is called a {\em simple key} if $K$ is a simple attribute,
otherwise it is called a {\em compound key}.\footnote{Note: We do not require compound
  keys to be minimal.} An argument position in an atom that
corresponds to a simple key attribute is referred to as a {\em keyed position}.

\begin{definition}~\label{def:chase}
  Given a conjunctive query $$Q=R_0(u_0) \leftarrow R_1(u_1) \wedge \ldots \wedge R_n(u_m),$$ we define
  $chase(Q)$ to be the result of iteratively performing the following replacements:
  \begin{itemize}
    \item{Given two atoms $R_i(u_j)$ and $R_i(u_k)$ of the same relation, with the $p^{th}$ position a key
        for relation $R_i$, if the variable at the $p^{th}$ position of $u_j$ is the same as the variable
        at the $p^{th}$ position of $u_k$, then for each $h \in 1,\ldots,|u_j|$ let $X$ be the variable
        that occurs at position $h$ in $u_j$.  We replace every instance of $X$ that occurs anywhere in
        the query by the variable occurring at position $h$ of $u_k$, and proceed with the updated
        $u_i$'s.  Finally, we remove the term $R_i(u_j)$ from the conjunctive query.}
  \end{itemize}
\end{definition}

While the above definition only applies to queries with simple keys, the chase operator extends to arbitrary functional dependencies, though we refer the reader to~\cite{chase2} for details.

The following fact confirms the intuition that the substitutions in Definition~\ref{def:chase} do not affect the result of the query.

\begin{fact}~\label{fact:chase}~\cite{chase2, ASUchase, chase3}
For any instance, the result of applying the query $chase(Q)$ is identical to the output of applying $Q$.
\end{fact}

\subsection{The Color Number}~\label{sec:coloring}

We restate the definitions from~\cite{GLV} of \emph{valid coloring} and the \emph{color number} $C(Q)$ of a query, and state the size bounds of~\cite{GLV}.

\begin{definition}~\label{def:valid_coloring2}
  Given a conjunctive query $$Q=R_0(u_0) \leftarrow R_1(u_1) \wedge
  \ldots \wedge R_n(u_m),$$ and the set of functional dependencies for
  each input relation, a \emph{valid coloring} of $Q$ with $c$ colors
  is a coloring $\mathcal{C}:var(Q) \rightarrow 2^{\{1,\ldots, c\}}$ assigning to each variable $X\in var(Q)$ a set of colors $\mathcal{L}(X) \subset \{1,\ldots, c\}$, consisting of zero or more colors such that the following
condition is satisfied:
  \begin{itemize}
% subseteq:
    \item{For each functional dependency $X_1,\ldots,X_k \rightarrow Y,$ $$\mathcal{L}(Y) \subseteq
        \bigcup_i \mathcal{L}(X_i).$$}
  \end{itemize}
\end{definition}

\begin{definition}~\label{def:color_num}
The \emph{color number} of a query $Q=R_0(u_0) \leftarrow R_1(u_1) \wedge \ldots \wedge R_n(u_m),$ denoted $C(Q)$, is the maximum over valid colorings of $Q$ of the ratio of the total number
of colors appearing in the output variables $u_0$, to the maximum number of colors appearing in any given
$u_i$, for $i \ge 1$.  Formally: $$C(Q):= \max_{\text{colorings }} \frac{|\bigcup_{X_j \in u_0}
\mathcal{L}(X_j)|}{\max_{i\ge 1} |\bigcup_{X_j \in u_i} \mathcal{L}(X_j) |}.$$
\end{definition}

The main theorem of~\cite{GLV} is that the color number yields a tight bound on the worst-case size increase of general conjunctive queries either without functional dependencies, or with a set of simple functional dependencies (or simple keys).  Formally, the following theorem is proven:

\begin{theorem1}[Theorem 4.7 from~\cite{GLV}]
Given a query $Q=R(u_0) \leftarrow R_1(u_1) \wedge \ldots \wedge
R_n(u_m)$ and set of simple functional dependencies, $$|Q(D)|
\le \rmax(Q,D)^{C\left(chase(Q)\right)}.$$ Furthermore, this bound is essentially tight: for any $N>0$, there exists a database $D$ with $\rmax(Q,D) \le rep(Q) \cdot N$, and $|Q(D)|= N^{C(Q)},$ where $rep(Q)$ is the maximum number of times any specific relation $R_i$ appears in $Q$.
\end{theorem1}

Additionally, it was shown that, in the setting in which general functional dependencies are given, the color number yields a lower bound.  Specifically,

\begin{prop1}[Proposition 6.3 from~\cite{GLV}]
Given a query $Q=R_0(u_0) \leftarrow R_1(u_1) \wedge \ldots \wedge R_n(u_m)$ and set of functional dependencies,
there exists an instance $D$ in which $$|Q(D)| \ge \left(\frac{\rmax(Q,D)}{rep(Q)} \right)^{C\left(chase(Q)\right)}.$$
\end{prop1}

The proof of the above proposition is via a construction.  This construction provides some insight into the relationship between the colorings of the variables, and conditional entropies, and we give a simplified proof in the case that $m=n$ in Appendix~\ref{construction}.

\subsection{Conditional Entropy and Information Measures}~\label{sec:entropy_defs}

In this section we state the basic definitions of \emph{conditional entropy} and \emph{information measures}, and then state some facts about Shannon and non-Shannon information inequalities, which will prove useful in the remainder of the paper.

\begin{definition}~\label{def:conditionalentropy}
For discrete random variables $X,Y$ with respective supports $\mathcal{X}, \mathcal{Y},$ the conditional entropy of $X$ given $Y$, denoted by $H(X|Y)$ is given by $$H(X|Y):=\sum_{y \in \mathcal{Y}} p(y)H(X|Y=y) = - \sum_{x \in \mathcal{X}} \sum_{y \in \mathcal{Y}} p(x,y) \log\left(p(x|y)\right).$$
\end{definition}

The following fact follows from the above definition:
\begin{fact}~\label{fact:entropy}
For discrete random variables $X,Y$ with respective supports $\mathcal{X}, \mathcal{Y},$ $$H(X,Y) = H(X)+H(Y|X).$$
\end{fact}

\begin{definition}\label{mutual-information}
For discrete random variables $X,Y$, as above, the \emph{mutual information} between $X$ and $Y$ is $$I(X;Y):=\sum_{x \in \mathcal{X},y \in \mathcal{Y}} p(x,y) \log \frac{p(x,y)}{p(x)p(y)}.$$
\end{definition}

The following fact follows from the above definition:
\begin{fact}~\label{fact:entropy1}
For discrete random variables $X,Y$ as above, $$I(X;Y)=I(Y;X)=H(X)+H(Y)-H(X,Y) = H(X)-H(X|Y).$$
\end{fact}

\begin{definition}
For discrete random variables $X_1,\ldots,X_n$ with respective supports $\mathcal{X}_1,\ldots,\mathcal{X}_n$, and $n\geq 3$, we recursively define their mutual information as $$I(X_1;\ldots;X_n)=I(X_1;\ldots;X_{n-1})-I(X_1;\ldots; X_{n-1}|X_n),$$ where the \emph{conditional mutual information} is defined as $$I(X_1;\ldots;X_{n-1}|X_n)=\sum_{x_n\in\mathcal{X}_n} p(x_n)(I(X_1;\ldots;X_{n-1})|X_n=x_n),$$ and where for $n=2$, mutual information is as defined in Definition \ref{mutual-information}.
\end{definition}

\begin{figure}
\begin{center}
\epsfig{file=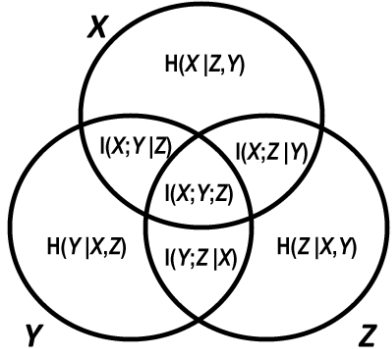,width=.42\textwidth} \caption{The generic information diagram of $X,Y,Z.$  Note that the set-theoretic properties of these information measures allows various information equalities to be read off from such a diagram; for example, $I(X;Y)=I(X;Y;Z)+I(X;Y|Z),$ and $H(Z)=I(X;Y;Z)+I(X;Z|Y)+I(Y;Z|X)+H(Z|X,Y).$}\label{fig:information_diagram}
\end{center}
\end{figure}

Unsurprisingly, the above information measures have a set-theoretic structure, and can be represented in an \emph{information diagram}, from which basic relations between information measures can be easily read off.  Figure~\ref{fig:information_diagram} illustrates a general information diagram for three variables.  The following facts follow from the previous definitions, and can easily be seen by considering the associated information diagram. (We refer the reader to Chapter 3 of ~\cite{yeung_book} for proofs of these facts and rigorous definition of the set-theoretic structure of information measures.)

\begin{fact}~\label{fact:entropy_set}
  For discrete random variables $X_1,\ldots,X_n$, and any disjoint sets $K,K' \subseteq [n],$:
  $$H(X_K|X_{K'}) = \sum_{S: S\cap K \neq \emptyset, S \cap K' = \emptyset} I(S|X_{[n]-S}),$$
   $$I(K|X_{K'}) = \sum_{S: S \supseteq K, S \cap K'=\emptyset} I(S| [n]-S),$$
   where $I(S|X_{S'})$ denotes $I(X_1; \ldots ; X_j|X_{S'}),$ for $S=[j]$.  Note that we avoid the notation $I(X_S|X_{S'}),$ which has the interpretation of $I(X_1,\ldots,X_j|X_{S'})=H(X_S|X_{S'}).$
\end{fact}

We now define the basic information inequalities.

\begin{definition}
For discrete random variables $X_1,\ldots,X_n$ as above, and for a subset $K\subset [n]$, denoting by $X_K$ the tuple of all $X_i$ for $i\in K$, the \emph{Shannon information inequalities} consist of all inequalities of the form $$H(X_i|X_{[n]-\{i\}})\geq 0,$$ for all $i\in [n]$, and $$I(X_i;X_j|X_K)\geq 0,$$ for all $i\neq j \in n$ and $K\subset [n]-\{i,j\}$.
\end{definition}

We note that, as above, the mutual information expressions can be reexpressed in terms of entropies.  For example, $I(X_i;X_j|X_K)=H(X_i|X_K)-H(X_i|X_j,X_K)=H(X_i,X_K)+H(X_j,X_K)-H(X_K)-H(X_i,X_j,X_K)$.
(See \cite{yeung_book}, Chapter 14 for further discussion of the Shannon inequalities.)

The Shannon information inequalities are well-understood and were, initially, hypothesized to essentially capture the space of valid entropy configurations.  However, in a breakthrough work in 1998, Zhang and Yeung showed that there are fundamental constraints on this space that are not captured by the Shannon inequalities, even for as few as four random variables~\cite{yeung98}.  This accounts for the lack of tightness in our upper bound.

\section{Size Bounds}~\label{sec:size}

We begin by giving our linear programming upper bound for the worst-case size increase.  Throughout this section, we admit a slight abuse of notation, and refer to the entropy of a set of attributes of a database, interpreted in the natural way: given a database table with attribute set $A=\{X_1,\ldots,X_k\}$, some fixed probability distribution $\mathcal{D}$ over the tuples of the table, and two subsets $S,S'\subseteq A$, we refer to the conditional entropy $H_{\mathcal{D}}(S|S')$ where $S,S'$ respectively are interpreted to be the discrete random variables whose possible values consist of the $|S|$, respectively $|S'|-$tuples of values that the corresponding variables have in the tuples of the database table, with probabilities given according to $\mathcal{D}$.

\begin{theorem}~\label{thm:upper_bound}
Given a query $Q=chase(Q)=R_0(u_0) \leftarrow R_1(u_1) \wedge \ldots \wedge R_n(u_m),$ with $var(Q)=\{X_1,\ldots,X_k\},$ and a set of arbitrary functional dependencies, for any database $D$, $$|Q(D)| \leq \text{rmax}(Q,D)^{s(Q)},$$ where $\text{rmax}(Q,D)$ is the size of the largest relation among $R_1,\ldots,R_n$ in $D$, and $s(Q)$ is the solution to the following linear program:
\begin{eqnarray*} \text{maximize } & h(u_0) & \\
    \text{subject to } & h(u_i) \le 1 & \forall i \ge 1 \\
    & h(x_t|x_{i_1},\ldots, x_{i_j}) = 0 &\text{for each f.d. } X_{i_1},\ldots,X_{i_j} \rightarrow X_t \\
    & h(x_i|x_{[k]-\{i\}}) \ge 0 & \forall i \in [k] \\
    & I(x_i;x_j|x_S) \ge 0 & \forall i,j \in [k] \text{ and } S \subseteq [k]-\{i,j\},
\end{eqnarray*}
where the variables of the linear program are the (unconditional) entropies $h(x_S)$ for all $S\subseteq [k]$, and the expressions involving mutual information or conditional entropies appearing in the constraints are implicitly considered to stand in for the corresponding linear expressions of these variables (as described in Section \ref{sec:entropy_defs}).
\end{theorem}
\begin{proof}
  The first step in the proof is to establish the connection between entropy and worst-case size increases.  Given our query $Q$ and database $D$, let $c$ be such that $|Q(D)| = \rmax(Q,D)^c.$  Let $Q'= R_0'(var(Q)) \leftarrow R_1(u_1) \wedge \ldots \wedge R_n(u_m)$ be the query derived from $Q$ by including all query variables in the output, and define the distribution $\mathcal{D}$ over the tuples of $Q'(D)$ to be such that the marginal distribution $\mathcal{D}_{u_0}$ over the values of the $|u_0|$-tuples corresponding to variables in $u_0$ is the uniform distribution.  Note that such a choice for $\mathcal{D}$ is not necessarily unique, unless $u_0=var(Q)$.  Let $H_{\mathcal{D}}(u_i)$ denote the entropy of the projection of the distribution $\mathcal{D}$ onto the positions labeled by the variables of $u_i$.  Observe that for any $i \in [m],$
\begin{eqnarray}\frac{H_{\mathcal{D}}(u_0)}{H_{\mathcal{D}}(u_i)} & \ge & \frac{H_{\mathcal{D}}(u_0)}{H_{unif_i}(u_i)} \ge \frac{\log(|Q(D)|)}{\log(|R_i(D)|)} \ge c, \end{eqnarray}
where $unif_i$ is the uniform distribution over the tuples of $R_i(D).$  This provides the motivation for the form of our linear program: maximizing the entropy of $u_0$ while bounding the entropies of each $u_i$.

To see that the value of the above linear program provides an upper bound on $\frac{\log(|Q(D)|)}{\log(|R_i(D)|)},$ note that for any set $S \subset [k]$, the quantity $\frac{H_{\mathcal{D}}(S)}{\max_{i\ge 1} H_{\mathcal{D}}(u_i)}$ satisfies all the constraints that the corresponding variable $h(S)$ is subject to in the linear program, including the last two sets of constraints that represent the Shannon information inequalities, and thus by Equation (2) the value of the solution to the linear program must be at least $\frac{\log(|Q(D)|)}{\log(|R_i(D)|)}.$
\end{proof}

In order to make the size bound given by the solution to the linear program of Theorem~\ref{thm:upper_bound} tight, we would need to add additional constraints so as to enforce the \emph{non-Shannon} information inequalities.  Unfortunately, it was recently shown that even for just four variables, there are infinitely many independent such inequalities~\cite{matus_inf}.

 We note that the jump in difficulty of establishing tight size bounds occurs when the left-hand sides of functional dependencies go from having single variables, to having 2 variables.  It is not hard to show that any size bounds for the case where functional dependencies have left-hand sides with at most two variables can be extended to work for arbitrary functional dependencies, via the following proposition.

  \begin{proposition}~\label{fact:twofds}
  Given a query $Q=chase(Q)$ and set of functional dependencies, there exists a query $Q'$ with the following properties:
  \begin{itemize}
    \item{each functional dependency of $Q'$ has at most two variables on its left-hand side,}
    \item{$Q'=chase(Q'),$}
    \item{the set of functional dependencies of $Q'$ is at most polynomially larger than that of $Q$,}
    \item{the description of $Q'$ is at most polynomially larger than that of $Q$,}
    \item{the worst-case size increase of $Q$ and $Q'$ are identical.}
    \item{$C(Q)=C(Q')$.}
  \end{itemize}
\end{proposition}
\begin{proof}
  We shall iteratively remove functional dependencies from $Q$ that have 3 or more variables occurring on their left-hand sides, via the addition of a (polynomial number) of additional variables, relations, and functional dependencies.

  Given a functional dependency $X_1 \ldots X_k \rightarrow Y,$ we add a relation $R(X_1 X_2 Z)$, with the new variable $Z$, together with the functional dependencies $X_1 X_2 \rightarrow Z, Z \rightarrow X_1, Z \rightarrow X_2.$  We then add the relation $R'(Z X_3 \ldots X_k Y),$ together with the functional dependency $Z X_3 \ldots X_k \rightarrow Y.$  Finally, we remove the functional dependency $X_1 \ldots X_k \rightarrow Y$ from the set of functional dependencies.

  Iteratively applying the above procedure until there are no more functional dependencies (other than implied ones) with more than two variables on their left-hand sides clearly results in a query $Q'$ with at most a polynomially longer description, and polynomially more functional dependencies.  Additionally, since all new relations are distinct, and all original functional dependencies are implied by the new set of functional dependencies, $chase(Q')=Q'.$  To see that the size increase of $Q'$ is the same as that of $Q$, note after each single iteration of the above procedure, the size increase must remain unchanged, as the values taken by variables $X_1,X_2$ dictate that taken by $Z$, and vice versa, defining a $1:1$ mapping between tuples of $Q(D)$ and tuples of the result of the query generated after one step of the procedure.  To conclude,  there is a natural mapping between valid colorings of $Q$, and the query obtained after one step of the above procedure, namely $\mathcal{L}(Z)\leftrightarrow \mathcal{L}(X_1) \cup \mathcal{L}(X_2).$
\end{proof}

\section{The Color Number and Entropy}~\label{sec:alt}

We now reexamine the color number in an effort to better understand the types of entropy structures that it can capture.  As the following proposition shows, the color number can be defined via the linear program of Theorem~\ref{thm:upper_bound} with the addition of some extra constraints on the entropies.  In particular, we require extra constraints that enforce that all mutual information measures be nonnegative.  (Note that the Shannon inequalities imply that all mutual information measures of two variables be nonnegative; however, as Figure~\ref{fig:entropy_diagram} depicts, the mutual information of more than two variables can be negative.)

\begin{theorem}
  Given a query $Q=chase(Q)=R_0(u_0) \leftarrow R_1(u_1) \wedge \ldots \wedge R_n(u_m),$ with $var(Q)=\{X_1,\ldots,X_k\},$ and a set of arbitrary functional dependencies, $C(Q)$ is equal to the solution to the following linear program:
\begin{eqnarray*} \text{maximize } & h(u_0) & \\
    \text{subject to } & h(u_i) \le 1 & \forall i \ge 1 \\
    & h(x_t|x_{i_1},\ldots, x_{i_j}) = 0 &\text{for each f.d. } X_{i_1},\ldots,X_{i_j} \rightarrow X_t \\
    & I(x_{i_1};,\ldots;x_{i_j}|x_{[k]-\{i_1,\ldots,i_j\}}) \ge 0 & \forall \text{ sets }\{i_1,\ldots,i_j\}=S \subseteq [k],
\end{eqnarray*}
where the variables of the linear program are the (unconditional) entropies $h(x_S)$ for all $S\subseteq [k]$, and the expressions involving mutual information or conditional entropies appearing in the constraints are implicitly considered to stand in for the corresponding linear expressions of these variables (as described in Section \ref{sec:entropy_defs}).
\end{theorem}
\begin{proof}
  We first show that given any valid coloring achieving color number $C(Q)$, we can find a feasible point for the linear program with value $C(Q)$.  Given a valid coloring in which at most $r$ colors occur together in the labels of any input atom, for every set $S \subseteq [k],$ we set $$I(S|x_{[k]-S}) = \frac{| \bigcap_{i \in S} \mathcal{L}(X_i)-\bigcup_{i \not \in S}\mathcal{L}(X_i)|}{r},$$ where $I(S|x_{[k]-S})$ denotes $I(x_{i_1};\ldots;x_{i_j}|x_{[k]-S}),$ with $S = \{X_{i_1},\ldots,X_{i_j}\}.$  Note that these $2^n$ mutual information values are sufficient to determine the values of all variables in the linear program.  In particular, these $2^n$ mutual information measures are the values that would appear in an information diagram.  From Fact~\ref{fact:entropy_set}, for any disjoint sets $T,T' \subseteq[k],$ we will now express $I(T|x_{T'})$ in terms of the color labels.  We note that for distinct sets $S_1,S_2$, the corresponding sets of labels $\bigcap_{i \in S_j} \mathcal{L}(X_i)-\bigcup_{i \not \in S_j}\mathcal{L}(X_i)$ will be disjoint, because these sets consist of exactly those colors appearing in the labels of each element of $S_j$ and not in any of the labels of elements not in $S_j$.  Thus the sum in Fact~\ref{fact:entropy_set} may be expressed in terms of the size of the union of these sets for $S$ containing $T$ and disjoint from $T'$.  It is straightforward to see that this union consists of exactly those colors appearing in the labels of each element of $T$ and not in any of the labels of elements of $T'$, yielding: $$I(T|x_{T'}) = \frac{| \bigcap_{i \in T} \mathcal{L}(X_i)-\bigcup_{i \in T'}\mathcal{L}(X_i)|}{r}.$$

  It is now easy to see that this construction yields a feasible point for the linear program. First observe that all the information inequalities are trivially satisfied, since for every set $S \subseteq [k],$ $I(S|x_{[k]-S}) \ge 0$ in our construction.  To see that the equality constraints given by the functional dependencies are observed, note that the dependency $X_1,\ldots,X_j \rightarrow X_{j+1}$ implies that $\mathcal{L}(X_{j+1})-\bigcup_{i \in [j]}\mathcal{L}(X_i) = \emptyset,$ and thus in the above assignment, $I(x_{j+1}|x_{[j]})=0,$ as desired.  (Note that, by definition, $h(x_{j+1}|x_{[j]})=I(x_{j+1}|x_{[j]}).$)  Finally, to see that the first set of constraints are observed, note that for any $j \le k,$ $h(x_{[j]})=\sum_{S \text{ s.t. } S \cap [j] \neq \emptyset} I(S|x_{[j]-S}),$ which, by our construction, is precisely $\frac{|\bigcup_{i \in [j]} \mathcal{L}(X_i)|}{r},$ which is bounded by 1 whenever $S$ is the index set of an input atom, and which will equal $C(Q)$ when $S$ is the index set of $u_0$ by the definition of the color number.

  For the other direction, given a rational feasible point for the linear program with objective function value $v,$ where all variables have values $r_i/q,$ for integers $r_i,q$, with $q$ being the common denominator, we will construct a coloring with color number $C(Q)$.  The final set of constraints of the LP implies that for any set $S \subseteq [k],$ $I(S|x_{[k]-S}) =\frac{r_{S}}{q} \ge 0.$  Furthermore, since our feasible point is rational, $r_{S} \in \N.$  To populate our coloring, we begin with the empty coloring, and then for each $S \subseteq [k],$ we add $q \cdot i(S|x_{[k]-S})$ unique colors to the labels of all $X_i$ for which $i \in S.$  To see that this coloring obeys the functional dependencies, note that for $X_1,\ldots,X_j \rightarrow X_{j+1},$ we have that $I(x_{j+1}|X_{[j]})=0,$ and thus by Fact~\ref{fact:entropy_set}, for any $S \subset [k]-[j]$ such that $j+1 \in S,$ $I(S|X_{[k]-S})=0$, from which it follows that in our construction $\mathcal{L}(X_{j+1}) \subseteq \bigcup_{i \in [j]} \mathcal{L}(X_i).$  Finally, to see that the color number is at least the value $v$, of the linear program, note that by Fact~\ref{fact:entropy_set}, a total of $$\sum_{S \subseteq [k] \text{ s.t. } S \cap K \neq \emptyset} q \cdot I(S|X_{[k]-S})=q \cdot h(X_S)$$ unique colors are assigned to each set $X_S$, and thus the color number is at least $h(u_0),$ as desired.
\end{proof}

\begin{remark}
  From the above characterization of the color number, it follows that for all the settings in which the color number yields a tight bound on the worst-case size increase (i.e. when no functional dependencies are specified, or only simple dependencies), there exist worst-case instances whose corresponding information diagrams have only nonnegative entries.
\end{remark}

\subsection{A Super-Constant Gap}

Leveraging the understanding of the entropy structures that are compatible with the color number given by the previous theorem, we now show that there is a super-constant gap between the exponent of the true worst-case size increase, and the color number (in the case of general functional dependencies).  We suspect, however, that in the majority of practical applications, this gap between the upper and lower bounds will be small.

\begin{theorem}
  For any fixed constant $\alpha \in \R,$ there exists a conjunctive query $Q$ and set of functional dependencies, and database $D$, such that $|Q(D)| > \rmax(Q,D)^{\alpha C(chase(Q))}.$
\end{theorem}
\begin{proof}
    We shall construct a family of queries, and associated databases whose color numbers fall short of the true size increase by a superconstant factor.\footnote{Our construction is a generalization of a construction suggested to us by Daniel Marx.}  Fix an even integer $k$, and consider the following query $Q$ over $k^2/2$ variables $X_{i,j}$, for $i \in \{1,\ldots,k\},$ and $j \in \{1,\ldots,k/2\}$: $$Q = R(X_{1,1},\ldots,X_{i,j},\ldots,X_{k,k/2}) \leftarrow \bigwedge_{i=1}^{k/2} R_i(X_{1,i},\ldots,X_{k,i}) \wedge \bigwedge_{i=1}^k T_{i}(X_{i,1},\ldots,X_{i,k/2}).$$  Additionally, for each $j \in \{1,\ldots, k/2\}$ we impose the following functional dependencies: given any set $S \subset \{X_{1,j},\ldots,X_{k,j}\},$ with $|S| \ge k/2,$ for any $i,$ $$S \rightarrow X_{i,j}.$$

\begin{figure}
\begin{center}
\epsfig{file=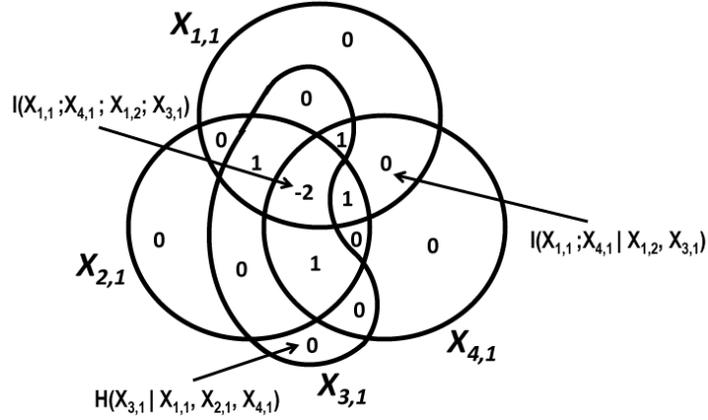,width=.74\textwidth} \caption{The information diagram of $X_{1,1},\ldots,X_{4,1}$ in our construction for $k=4.$  Note that any set of size 2 or more contains all the entropy of all four variables.  The negative mutual information $I(X_{1,1};X_{1,2};X_{1,3};X_{1,4})=-2$ suggests that no valid coloring can closely approximate the entropy structure, which is leveraged in our construction to yield a super-constant gap between the color number and worst-case size increase.}\label{fig:entropy_diagram}
\end{center}
\end{figure}

    Intuitively, the above construction has $k/2$ groups of $k$ variables, such that amongst any group, any set of $k/2$ of those variables suffice to recover the remaining $k/2$ variables in that group.  The information diagram of one group of the construction in the case $k=4$ is depicted in Figure~\ref{fig:entropy_diagram}.  Given any integer $N$, we will construct a database $D$ such that for all $i \in [k/2],j \in [k],$ we have $|R_i(D)| = N^{k/2}=|T_j(D)|$.  The values assigned to positions labeled by $X_{i,j}$ and $X_{i',j'}$ will be disjoint whenever $j \neq j'$; i.e. the values assigned each of the $k/2$ groups are disjoint.  Each of the $N^{k/2}$ tuples of $R_i(D)$ will be constructed so as to be Shamir $(k/2,k)$ secret shares~\cite{shamir}.  That is, given the values of any $k/2$ attributes $X_{1,i},\ldots,X_{k/2,i},$  the values of the remaining $k/2$ attributes can be uniquely determined, and for $S \subset \{X_{1,i},\ldots,X_{k,i}\},$ $$|\pi_S(R_i(D))|= \left\{ \begin{array}{cc} N^{|S|} & \text{if }|S| \le k/2, \\
    N^{k/2} & \text{if } |S| \ge k/2. \end{array} \right.$$

    Since $Q(D)$ consists of the complete join of each $R_i$, $|Q(D)| = \left(N^{k/2} \right)^{k/2} = N^{k^2/4},$ whereas the size of the largest input relation is $\rmax(Q,D) = N^{k/2}.$  We now show that $C(chase(Q))=C(Q) \le 2,$ which will complete our proof of the theorem.

    First observe that it suffices to consider the case that for $j \neq j',$ $\mathcal{L}(X_{i,j}) \cap \mathcal{L}(X_{i',j'}) = \emptyset,$ because, assuming otherwise, if the common color $c$ lay in the intersection, by removing the color $c$ from the labels $\mathcal{L}(X_{i'',j})$ for all $i'',$ we still have a valid coloring (since there are no functional dependencies between groups), and the color number could only have increased.  Let $r_i = |\bigcup_{j=1}^k \mathcal{L}(X_{j,i})|,$ and $t_i=|\bigcup_{j=1}^{k/2} \mathcal{L}(X_{i,j})|=\sum_{j=1}^{k/2} |\mathcal{L}(X_{i,j})|$ denote the number of colors assigned to the variables of each input atom.  Thus in any optimal coloring, we have $$|\bigcup_{X_{i,j}} \mathcal{L}(X_{i,j})|= \sum_{i=1}^{k/2} |\bigcup_{j=1}^k \mathcal{L}(X_{j,i})|=\sum_{i=1}^{k/2} r_i.$$
     Next, observe that each element of $\mathcal{L}(X_{i,j}),$ must occur in the labels of at least $k/2$ other variables $X_{i',j};$  if this were not the case, then there would exist a set $S \subset \{X_{1,j},\ldots,X_{k,j}\}$ of size $|S| \ge k/2,$ such that $\mathcal{L}(X_{i,j}) \not \subseteq \bigcup_{X_{i',j} \in S} \mathcal{L}(X_{i',j}),$ which violates one of the functional dependencies.  Thus it follows that $$\sum_{i=1}^k | \mathcal{L}(X_{i,j})| \ge \frac{k}{2} r_j.$$  To conclude, putting the above equations together, we have $$\sum_{i=1}^k t_i =\sum_{X_{i,j}}|\mathcal{L}(X_{i,j})| \ge \frac{k}{2} \sum_{i=1}^{k/2}r_i,$$ and thus there must be at least one $i$ such that $t_i \ge \frac{(k/2)\sum_{i=1}^{k/2} r_i}{k}=\frac{1}{2}\sum_{i=1}^{k/2}r_i,$ and thus $C(Q) \le 2.$
    \end{proof}

\section{Complexity Considerations}~\label{sec:complexity}

From a complexity standpoint, the results of the previous setting are not encouraging.  Both the upper bound, and lower bound of $C(Q)$ are given as the solutions to exponential-sized linear programs.    This prompts the question of whether one can efficiently determine anything about the size of the result, in this setting with general functional dependencies.  (It is shown in~\cite{GLV} that when one only has simple functional dependencies, tight size bounds can be efficiently computed.)  With general functional dependencies, even computing $chase(Q)$ can be intractable.  Nevertheless, we show that when $chase(Q)$ is given, or can be efficiently computed (for example, when all the input relations have bounded arities), we can efficiently decide whether the result of the query with a set of general functional dependencies can be any larger than the input relations. The proof relies on a proposition from~\cite{GLV}, and then reduces the question at hand to the satisfiability of a sequence of tractable SAT instances---one for each input relation.

\begin{theorem}~\label{thm:complexity}
  Given a conjunctive query $Q=R_0(u_0) \leftarrow R_1(u_1) \wedge \ldots \wedge R_n(u_m)$ with an arbitrary set of functional dependencies, such that $Q=chase(Q)$, it can be efficiently decided whether the results of $Q$ can be larger than the input relations, in which case there exists an instance $D$ with $|Q(D)| \ge \left(\frac{\rmax(Q,D)}{rep(Q)} \right)^{\frac{m}{m-1}}$.
\end{theorem}

The proof of the theorem relies on the following proposition:

 \begin{prop1}[Proposition 6.1 from~\cite{GLV}]~\label{thm:char}
  A query $Q=R_0(u_0) \leftarrow R_1(u_1) \wedge \ldots \wedge R_n(u_m)$ with arbitrary functional dependencies
  is sparsity preserving if, and only if $C\left(chase(Q)\right) = 1$.  Equivalently, for any database $D$, $|Q(D)| \le \rmax(Q,D)$ if, and only if $C\left(chase(Q)\right) = 1$.
 Furthermore, if $C(chase(Q))>1,$ then $C(chase(Q)) \ge \frac{m}{m-1}.$
 \end{prop1}

\begin{prevproof}{Theorem}{thm:complexity}
  By the above proposition, it suffices to show that one can decide whether $C(Q)>1$ in polynomial time.  First observe that a necessary and sufficient condition for $C(Q)>1$ is the existence of some coloring $\mathcal{C}$ such that for each relation $R_i$,with $i \ge 1$, there is a color $c_i$ such that $c_i \in \bigcup_{X_j \in u_0} \mathcal{L}(X_j),$ but $c_i \not \in \bigcup_{X_j \in u_i} \mathcal{L}(X_j)$.  We will represent this condition as a set of $n$ tractable SAT expressions, one for each input relation, as follows.  Our set of SAT variables will be $\{x_1,\ldots,x_{|var(Q)|} \},$ in natural correspondence with the set of query variables $V=\{X_1,\ldots,X_{|var(Q)|} \}.$

   From Proposition~\ref{fact:twofds} it suffices to prove our theorem in the case that all functional dependencies have at most two variables on their left-hand sides.    Given $p$ functional dependencies $X_{j_1} X_{k_1} \rightarrow X_{m_1}, \ldots, X_{j_p} X_{k_p} \rightarrow X_{m_p},$ our SAT expression for relation $i$ will have the form $$SAT_i=\bigwedge_{X_j \in u_i} \neg{x_j} \wedge \left( \bigvee_{X_j \in u_0} x_j \right) \wedge (x_{j_1} \vee x_{k_1} \vee \neg{x_{m_1}}) \wedge \ldots \wedge (x_{j_p} \vee x_{k_p} \vee \neg{x_{m_p}}).$$  Any satisfying assignment of $SAT_i$ yields a valid coloring of $Q$ that uses exactly 1 color, and has the property that no variable in $u_i$ has a color, but at least one variable in $u_0$ has a color; such a coloring is given by assigning all variables that are set to $false$ to not have the color, and all variables set to $true$ to have the color.  To see this, note that the first part of $SAT_i$ ensures that no variable occurring in $u_i$ can be $true$ in a satisfying assignment; the second part of $SAT_i$ ensures that at least one variable in the output projection will be colored, and the third part of $SAT_i$ ensures that the functional dependencies are respected.  Since any set of valid colorings can be combined to yield a valid coloring (by letting $\mathcal{L}_{1,2}(X_i)=\mathcal{L}_{1}(X_i) \cup \mathcal{L}_{2}(X_i)$), it follows that if, for all $i=1, \ldots, n$, $SAT_i$ is satisfiable, then there exists a coloring with $n$ colors, yielding $C(Q) \ge \frac{n}{n-1} > 1.$  Conversely, if, for some $i$, $SAT_i$ is not satisfiable, then there is no valid coloring of the variables in which some color appears in the output projection but not in the coloring of a variable of $u_i$, in which case $C(Q)=1.$

  What remains is to verify that $SAT_i$ can be solved efficiently.  We start by decomposing $SAT_i$ into its three basic components: $SAT_i = C_1 \wedge C_2 \wedge C_3,$ where $C_1 = \bigwedge_{X_j \in u_i} \neg x_j,$ $C_2 = \bigvee_{X_j \in u_0} x_j,$ and $C_3 = \bigwedge_{h=1,\ldots,p} (x_{j_h} \vee x_{k_h} \vee \neg x_{m_h}).$  We start by removing all variables $x_i$ from $C_2$ that appear negated in $C_1$.  Then, we simplify $SAT_i$ via a series of at most $|V|$ `passes'.  In each pass, we traverse each clause $(x_{j_h} \vee x_{k_h} \vee \neg x_{m_h})$ of $C_3$; if  $x_{m_h}$ occurs in $C_1$, then we remove the clause $(x_{j_h} \vee x_{k_h} \vee \neg x_{m_h})$ from $C_3$ and proceed.  Otherwise, if either $x_{j_h},$ or $x_{k_h}$ occur in $C_1$, we remove the occurring variable(s) from this clause in $C_3$ and proceed.  Finally, if a clause of $C_3$ consists of a single negated literal $\neg x_{\cdot},$ we remove that clause from $C_3$, and add the literal to $C_1$.  If no new variable is added to $C_1$ during a pass, this means that no additional passes will alter the clauses, so we halt.

  It is not hard to see that each pass does not alter the satisfiability of the expression $C_1 \wedge C_2 \wedge C_3$.  Furthermore, since each pass either adds at least one variable to $C_1$, or is the last pass, there will be at most $|V|$ passes.  If at any point a clause in $C_3$ becomes a single literal $x_i$ that also occurs in $C_1$, or $C_2$ consists of a subset of the variables occurring in $C_1$, then $SAT_i$ is clearly not satisfiable; if this does not occur, then no additional passes will alter the clauses, and a satisfying assignment for $SAT_i$ is given by setting all the variables in $C_1$ to be $false$, and all other variables to be $true$.
\end{prevproof}

\section{Conclusions}~\label{sec:conclusions}

We view the main contribution of this work as establishing a firm connection between worst-case size bounds and multivariate entropy structures, allowing the tools of information theory to be leveraged towards database analysis.  This connection promotes two main lines of future work.  The first direction is investigating whether one can explicitly characterize the worst-case size increase, even if that characterization is exponentially large.  It is also conceivable that, while exactly characterizing the size increase might not be possible, one can explicitly (and possibly even efficiently) compute an approximation of the worst-case size increase.  This seems like a deep and challenging question, and such a result would likely involve a significant advance in the understanding of the structure of non-Shannon type information inequalities.

The second direction is investigating which types of entropy structures arise from databases and their associated queries in practice.  Such an investigation would help determine where practical instances lie on the spectrum between the basic color number bounds and the more intricate bounds of Theorem~\ref{thm:upper_bound}.  Such database measures as sparsity and treewidth were introduced with corresponding goals in mind, and have proved effective at succinctly capturing the ease with which certain database operations can be done.  We propose the following measure of the entropy structure of a database and associated query, in the hope that it will succinctly capture this new facet of database complexity, as suggested by the results of this paper:

\begin{definition}
  \emph{The} knitted complexity \emph{of a database with respect to a query is the ratio of the sum of the absolute values of the mutual informations of all subsets of the query variables, to the sum of the (signed) mutual informations of all subsets of the query variables.}
\end{definition}

\section*{Acknowledgments}
We are deeply grateful to Daniel Marx, who first pointed out to us that the color number does not provide an upper bound on the worst-case size increase in the setting with general functional dependencies.

\bibliographystyle{abbrv}
\bibliography{size_bds_VV_bib}

\begin{thebibliography}{10}

\bibitem{abitebook}
S.~Abiteboul, R.~Hull, and V.~Vianu.
\newblock {\em Foundations of Data\-bases}.
\newblock Addison-Wesley, 1995.

\bibitem{ASUchase}
A.~V. Aho, Y.~Sagiv, and J.~D. Ullman.
\newblock Equivalence of relational expressions.
\newblock {\em SIAM J. of Computing}, 8(2):218--246, May 1979.

\bibitem{AGM}
A.~Atserias, M.~Grohe, and D.~Marx.
\newblock Size bounds and query plans for relational joins.
\newblock In {\em IEEE FOCS'08}, 2008.

\bibitem{chase3}
C.~Beeri and M.~Y. Vardi.
\newblock A proof procedure for data dependencies.
\newblock {\em J. ACM}, 31(4):718--741, 1984.

\bibitem{chandraMerlin}
A.~K. Chandra and P.~M. Merlin.
\newblock Optimal implementation of conjunctive queries in relational data
  bases.
\newblock In {\em ACM STOC}, 1977.

\bibitem{chaudhuri-qo}
S.~Chaudhuri.
\newblock An overview of query optimization in relational systems.
\newblock In {\em PODS 1998}.

\bibitem{doug07}
R.~Dougherty, C.~Freiling, and K.~Zeger.
\newblock Networks, matroids, and non-shannon information inequalities.
\newblock {\em IEEE Transactions on Information Theory}, 53(6):1949--1969,
  2007.

\bibitem{fagin03data}
R.~Fagin, P.~G. Kolaitis, R.~J. Miller, and L.~Popa.
\newblock Data exchange: Semantics and query answering.
\newblock In {\em ICDT}, 2003.

\bibitem{GLV}
G.~Gottlob, S.~T. Lee, and G.~J. Valiant.
\newblock Size and treewidth bounds for conjunctive queries.
\newblock In {\em PODS 2009}.

\bibitem{MM}
M.~Grohe and D.~Marx.
\newblock Constraint solving via fractional edge covers.
\newblock In {\em SODA 2006}.

\bibitem{Haas}
P.~J. Haas, J.~F. Naughton, S.~Seshadri, and A.~N. Swami.
\newblock Selectivity and cost estimation for joins based on random sampling.
\newblock {\em J. Comput. Syst. Sci.}, 52(3):550--569, 1996.

\bibitem{jarke-koch}
M.~Jarke and J.~Koch.
\newblock Query optimization in database systems.
\newblock {\em ACM Comput. Surv.}, 16(2):111--152, 1984.

\bibitem{kola05schema}
P.~Kolaitis.
\newblock Schema mappings, data exchange, and metadata management.
\newblock In {\em PODS}, 2005.

\bibitem{lenzerini}
M.~Lenzerini.
\newblock Data integration: a theoretical perspective.
\newblock In {\em PODS}, 2002.

\bibitem{LMS}
A.~Y. Levy, A.~O. Mendelzon, and Y.~Sagiv.
\newblock Answering queries using views.
\newblock In {\em PODS 1995}.

\bibitem{chase2}
D.~Maier, A.~O. Mendelzon, and Y.~Sagiv.
\newblock Testing implications of data dependencies.
\newblock {\em ACM Trans. Database Syst.}, 4(4):455--469, 1979.

\bibitem{matus_inf}
F.~Mat\'u\v{s}.
\newblock Infinitely many information inequalities.
\newblock In {\em 2007 IEEE International Symposium on Information Theory},
  Nice, France, 2007.

\bibitem{matus07}
F.~Mat\'u\v{s}.
\newblock Two constructions on limits of entropy functions.
\newblock {\em IEEE Transactions on Information Theory}, 53(1):320--330, 2007.

\bibitem{OlkenR90}
F.~Olken and D.~Rotem.
\newblock Random sampling from database files: A survey.
\newblock In {\em Proc. of Stat. and Scientific Database Management}, 1990.

\bibitem{pip86}
N.~Pippenger.
\newblock What are the laws of information theory?
\newblock In {\em 1986 Special Problems on Communication and Computation
  Conference}, Palo Alto, CA, 1986.

\bibitem{shamir}
A.~Shamir.
\newblock How to share a secret.
\newblock {\em Commun. ACM}, 22(11):612--613, 1979.

\bibitem{swami94}
A.~N. Swami and K.~B. Schiefer.
\newblock On the estimation of join result sizes.
\newblock In {\em Advances in Database Technology - EDBT'94. 4th Int. Conf. on
  Extending Database Technology}, 1994.

\bibitem{yeung_book}
R.~W. Yeung.
\newblock {\em Information Theory and Network Coding}.
\newblock Springer Publishing Company, Incorporated, 2008.

\bibitem{yeung97}
Z.~Zhang and R.~W. Yeung.
\newblock A non-shannon-type conditional inequality of information quantities.
\newblock {\em IEEE Transactions on Information Theory}, 43(6):1982--1986,
  1997.

\bibitem{yeung98}
Z.~Zhang and R.~W. Yeung.
\newblock On characterization of entropy function via information inequalities.
\newblock {\em IEEE Transactions on Information Theory}, 44(4):1440--1452,
  1998.

\end{thebibliography}

\appendix

\section{Simplified Proof of Proposition 6.3 from~\cite{GLV}}~\label{construction}

For clarity, we state and prove the proposition in the case that each input relation occurs only once in the query, and thus $Q=chase(Q).$

\begin{proposition}
Given a query $Q=R_0(u_0) \leftarrow R_1(u_1) \wedge \ldots \wedge R_n(u_n)$ and set of functional dependencies,
there exists an instance $D$ in which $$|Q(D)| \ge \left(\rmax(Q,D) \right)^{C(Q)}.$$
\end{proposition}
\begin{proof}
  Given an integer $N$, and any valid coloring with $d$ colors, with $d'\le d$ colors appearing in the labels of the output variables, such that the coloring achieves color number $C(Q)$, we shall construct an instance of $D$ with the property that $|Q(D)| = N^{d'},$ and $\text{rmax}(Q,D) \le N^{d'/C(Q)}.$

Consider a table of arity $d$, with attributes $C_1,\ldots,C_d,$ corresponding to each of the $d$ colors.  We construct the table $T$ to have $N^d$ tuples, such that the projection $\pi_{C_{i_1},\ldots,C_{i_k}}(D)$ of $D$ onto any $k$ attributes $C_{i_1},\ldots,C_{i_k}$ has size $N^k$.  We denote the $N$ values that a given attribute $C_i$ may take by the values $i_1,\ldots,i_N.$   (Thus $T$ is just the total join of the $d$ columns of size $N$.)

Next, we populate a given relation $R_j$, that has variables $X_1,\ldots,X_k$ in the corresponding atom $u_j$.  Assume, without loss of generality that in the given coloring of $Q$, $\bigcup_{i=1,\ldots,k} \mathcal{L}(X_i) = \{1,\ldots, q\}.$  We populate $R_j$ with $N^q$ tuples derived from the $N^q$ tuples in $\pi_{C_1,\ldots,C_q}(T),$ where the values that attribute $X_i$ takes are given by an ordered list of the values taken by the $C_i's$ that are in $\mathcal{L}(X_i).$  To illustrate, say $q=3,$ and $(1_\cdot,2_{\cdot},3_{\cdot})$ is a tuple of $\pi_{C_1,\ldots,C_q}(T)$, if $R_j(XY)$ appears in $Q$, and $\mathcal{L}(X)=\{1,2\},\mathcal{L}(X)=\{2,3\},$ then we add the tuple $([1_{\cdot},2_{\cdot}],[2_{\cdot},3_{\cdot}])$ to $R_j,$ with the value $[1_{\cdot},2_{\cdot}]$ appearing in the first attribute of $R_j$.  From the definition of valid coloring, it follows that the constructed database satisfies all functional dependencies.  Additionally, by construction, if all variables appeared in the output, all $N^d$ tuples would appear in the output, and thus $|Q(D)|=N^{d'}.$  For each input relation $R_i,$ we have $|R_i(D)| =N^k,$ where $k=|\bigcup_{X \in u_i} \mathcal{L}(X)|$, as desired.
\end{proof}
\end{document}